\documentstyle{article}
\title{RADIATIVE COOLING FLOWS OF SELF-GRAVITATING FILAMENTARY CLOUDS}

\author{MOHSEN SHADMEHRI\thanks{shad@science1.um.ac.ir}\\   AND   \\JAMSHID GHANBARI\thanks{ghanbari@science1.um.ac.ir}\\\it Department of Physics, School of Sciences, Ferdowsi University, Mashhad, Iran}
\date{Accepted {\it by The Astrophysics and Space Science}}
\begin{document}
\font\eightrm=cmb10 at 8pt

\maketitle

\begin{abstract}
\font\eightrm=cmb10
\baselineskip=1\baselineskip
We study the dynamics of a self-gravitating cooling filamentary cloud using a simplified model. We concentrate on the radial distribution and restrict ourselves to quasi-hydrostatic, cylindrically symmetric cooling flows. For a power-law dependence of cooling function on the temperature, self-similar solutions which describe quasi-hydrostatic cooling flows are derived. We consider obtically thin filaments with a constant mass per unit length and the solutions are parameterized by their line masses. There is no polytropic relation between the density and the pressure. The filament experiences radiative condensation, irrespective of the $\gamma$, the gas specific heat ratio. So, the filament becomes denser due to the quasi-hydrostatic flows and the density at the center $(\rho_{c})$ increases in proportion to $(t_{0}-t)^{-1}$, where $t$ denotes the time. The term, $t_{0}$, denotes an epoch at which the central density increases infinitely. We also found that the radius of the filament $(r_{c})$ decreases in proportion to $(t_{0}-t)^{\frac{1}{2}}$.
\\
\\{\it Subject headings}: hydrodynamic - instabilities - ISM: clouds - stars: formation
\end{abstract}

\section{INTRODUCTION}
\baselineskip=1\baselineskip
Interstellar molecular clouds are very complex structures in which many physical processes come into play. In these clouds, rich and complex structures have been seen (see, e.g., Myers 1991). Since most star formation essentially occurs in molecular clouds, understanding the origin and evolution of their structures is very important.

Filamentary structures associated with clumps and cores are very common (e.g., Schneider \&\ Elmegreen 1979; Houlahan \&\ Scalo 1992; Wiseman \&\ Adams 1994; Alves et al. 1998; Harjunpaa et al. 1999). It is now widely accepted that this hierarchical structure has been formed by gravitational instability. In this scenario for the formation of the hierarchical structure, four distinct phases have been studied: (a) gravitational instability along the filamentary cloud leads to the formation of clumps (e.g., Nagasawa 1987; Nakamura, Hanawa, \&\ Nakano 1993; Gehman, Adams, \&\ Watkins 1996; Nagai, Inutsuka, \&\ Miyama 1998; Fiege \&\ Pudritz 2000); (b) these clumps become geometrically thin disks, perpendicular to the filament axis (Nakamura, Hanawa, \&\ Nakano 1995; Nakamura et al. 1999); (c) The disks collapse toward their centers and, because of gravitational instability, each of them fragments into cloud cores (Nakamura \&\ Hanawa 1997; Saigo \&\ Hanawa 1998); and finally (d) these cores are scattered by gravitational interaction.

It must be noted that in this scenario, the overall collapse of a filamentary cloud is slower than the growth of local density fluctuations. In fact, it has been assumed that the parent filamentary cloud is well below the critical mass per unit length required for radial collapse to a spindle. There is a possibility that the filamentary cloud collapses, however, and several authors have studied the collapse of filaments (Bonnell \&\ Bastien 1991; Inutsuka \&\ Miyama 1993; Kawachi \&\ Hanawa 1998).

Most of the previous works have studied the collapse of spherically symmetric clouds (e.g., Larson 1972; Terebey, Shu, \&\ Cassen 1984; Foster \&\ Chevalier 1993). Calculations showed that the temperature of a collapsing spherical molecular cloud remains roughly constant in the early phases of star formation (see, e.g., Larson 1969). Investigating the collapse of a spherical cloud using three dimensional numerical simulations is very difficult. But there exists asymptotic self-similar solutions for the collapse of a spherical cloud (Larson 1969; Penston 1969; Shu 1977). Also, self-similar solutions for the collapse of a filamentary cloud were investigated, and different sets have been found (Inutsuka \&\ Miyama 1992; Kawachi \&\ Hanawa 1998; Semelin, Sanchez, \&\ de Vega 1999). Kawachi \&\ Hanawa (1998) investigated the gravitational collapse of a filamentary cloud using zooming coordinates (Bouquet et al. 1985). They used a polytropic equation of state to indicate the effects of deviations from isothermality in the collapse.

In fact, all of these authors followed the usual (Jeans) scenario for star formation. But as we shall show in this paper, there exists another (non-Jeans) scenario for star formation in filamentary clouds. Because of radiative cooling, a filamentary cloud which is in equilibrium, may evolve over a longer period of time. Thus, radiative condensation  flows can be formed in a cooling filamentary cloud. So, although the cloud is in a quasi-hydrostatic state, we may have density collapse, due to the presence of a radiative condensation flow. The same problem has been investigated for the dynamics of cooling spherical clouds (Meerson, Megged, \&\ Tajima 1996; hereafter MMT). In the present work, we have studied radiative cooling flows in a filamentary cloud, using a family of similarity solutions and instead of a polytropic equation of state, the energy equation is used. We found solutions that did not indicate any polytropic relation between  pressure and density. Since we have considered the constraint of a constant mass per unit length, the solutions were parameterized by their line masses. We show that for the power-law radiative cooling function, the dynamics of quasi-hydrostatic, cylindrically symmetric flows in a filamentary cloud is independent of the specific heat ratio of the gas.

The plan of this paper is as follows. We present the general mathematical framework and equations in $\S$ 2. By introducing similarity solutions, the dynamics of a radiative cooling filamentary cloud has been studied in $\S$ 3. We concentrate on quasi-hydrostatic flows. In $\S$ 4, after discussing the behaviour of central density and the radius of the cloud, we conclude with the implications of our model.
\section{GENERAL FORMULATION}
We present here the basic equations used to describe cylindrically symmetric radiative cooling flows. In this work, we neglect the effects of viscosity and magnetic fields. In cylindrical coordinates $(r, \varphi, z)$, which the filamentary cloud is assumed to be long in the $z-$direction, we have:
\begin{equation}
\frac{\partial\rho}{\partial t}+\frac{1}{r}\frac{\partial}{\partial r}(r\rho v)=0,
\end{equation}
\begin{equation}
\rho(\frac{\partial v}{\partial t}+v\frac{\partial v}{\partial r})+\frac{\partial p}{\partial r}+\rho\frac{\partial \Psi}{\partial r}=0 ,
\end{equation}
\begin{equation}
\frac{1}{r}\frac{\partial}{\partial r}(r\frac{\partial\Psi}{\partial r})=4\pi G\rho ,
\end{equation}
where $\rho, v, p, \Psi$ denote the gas density, radial velocity, pressure, and gravitational potential, respectively. We write the energy equation and the equation of state as follows:
\begin{equation}
\frac{1}{\gamma -1}(\frac{\partial p}{\partial t}+v\frac{\partial p}{\partial r})+\frac{\gamma}{\gamma -1}\frac{p}{r}\frac{\partial}{\partial r}(rv)+ \Lambda(\rho, T)=0,
\end{equation}
\begin{equation}
\\p=\frac{R_{g}}{\mu_{g}}\rho T,
\end{equation}
where $\Lambda(\rho, T)$ is the radiative loss function and $R_{g}$ and $\mu_{g}$ are the gas constant and the effective molar mass, respectively. Neglecting any external heating, we can fit the radiative loss function by a power law,
\begin{equation}
\Lambda(\rho, T)= A_{\nu}\rho^{2} T^{\nu},
\end{equation}
where $\nu$ and $A_{\nu}$ are constants. Depending on selected intervals of temperature, these parameters can be determined (Spitzer 1978).

Our scope of interest is quasi-hydrostatic flows in filamentary clouds. In fact, this situation occurs when cooling is slow on the dynamic time scale. For quasi-hydrostatic flows, the equation (2) becomes
\begin{equation}
\frac{\partial p}{\partial r}+\rho \frac{\partial \Psi}{\partial r}\simeq 0.
\end{equation}
The same relation has been used by MMT.

To simplify the problem, we introduce the dimensionless variables according to
\begin{equation}
\rho\rightarrow\hat{\rho}\rho, p\rightarrow\hat{p}p, \Psi\rightarrow\hat{\Psi}\Psi, T\rightarrow\hat{T}T, v\rightarrow\hat{v}v, r\rightarrow\hat{r}r, t\rightarrow\hat{t} t,
\end{equation}
where
\begin{equation}
\hat{\rho}=\rho_{0}, \hat{p}=p_{0}, \hat{T}=T_{0}, \hat{\Psi}=\frac{p_{0}}{\rho_{0}}, \hat{r}=(\frac{p_{0}}{4\pi G\rho_{0}^2})^\frac{1}{2}, \hat{t}=\frac{p_{0}}{(\gamma-1)A_{\nu}\rho_{0}^2 T_{0}^{\nu}}, \hat{v}=\frac{\hat{r}}{\hat{t}}.
\end{equation}
Under the transformation equations (8), the equations (1) and (7) do not change and the rest of the equations are cast into these forms:
\begin{equation}
\frac{1}{r}\frac{\partial}{\partial r}(r\frac{\partial\Psi}{\partial r})=\rho ,
\end{equation}
\begin{equation}
\frac{\partial p}{\partial t}+v\frac{\partial p}{\partial r}+\gamma \frac{p}{r}\frac{\partial}{\partial r}(rv)+ \rho^{2-\nu} p^{\nu}=0.
\end{equation}
\section{SELF-SIMILAR SOLUTIONS}
It is useful to study the similarity solutions of equations (1), (7), (10) and (11). We introduce the similarity variable as $y=\frac{r}{r_{0}(t)}$, where $r_{0}(t)=(t_{0}-t)^{\frac{1}{2}}$ and $t<t_{0}$. Using this transformation, we have the following forms of $\rho$, $p$, $v$ and $\Psi$:
\begin{equation}
\rho(r,t)=(t_{0}-t)^{-1}R(y),
\end{equation}
\begin{equation}
\\p(r,t)=(t_{0}-t)^{-1}P(y),
\end{equation}
\begin{equation}
\\v(r,t)=(t_{0}-t)^{-\frac{1}{2}}V(y),
\end{equation}
\begin{equation}
\Psi(r,t)=S(y),
\end{equation}
where $R(y)$, $P(y)$, $V(y)$ and $S(y)$ satisfy these equations:
\begin{equation}
\\R+\frac{y}{2}\frac{dR}{dy}+\frac{1}{y}\frac{d}{dy}(yRV)=0,
\end{equation}
\begin{equation}
\frac{dP}{dy}+R\frac{dS}{dy}=0,
\end{equation}
\begin{equation}
\frac{1}{y}\frac{d}{dy}(y\frac{dS}{dy})=R,
\end{equation}
\begin{equation}
\\P+(\frac{y}{2}+V)\frac{dP}{dy}+\gamma \frac{P}{y}\frac{d}{dy}(yV)+R^{2-\nu}P^{\nu}=0.
\end{equation}

This form of similarity solutions deals with blow-up, or at least rapid growth, in a finite time $t_{0}$. Since we are interested in cooling flows of an isolated filamentary cloud, the constraint of a constant mass per unit length can be written as
\begin{equation}
\int_{0}^{\infty} \rho r dr=m=const,
\end{equation}
where $m=\frac{2G\rho_{0}}{p_{0}}M$ and $M$ is mass per unit length.

Now, we have a set of ordinary differential equations. It is interesting that equations (16), (17), (18) and (19) are integrable. Using equation (16) we obtain
\begin{equation}
\\yRV+\frac{1}{2}y^{2}R=c,
\end{equation}
where $c$ is a constant. The velocity of the flows on the axis of a filamentary cloud is assumed to be zero. So we obtain
\begin{equation}
\\V(y)=-\frac{y}{2}.
\end{equation}
It shows that the velocity of the cylindrically symmetric cooling flow is independent of $\nu$, the exponent of the power-law cooling function. MMT showed that for spherically symmetric cooling flows, we may have gas inflow or outflow, depending on the value of $\nu$ . Substituting equation (22) into (19), we find
\begin{equation}
\\P(y)=(\gamma-1)^{\alpha}[R(y)]^{\beta},
\end{equation}
where $\alpha=\frac{1}{\nu-1}$ and $\beta=\frac{\nu-2}{\nu-1}$. This relation shows that $\nu$ may have any positive or negative values except $\nu=1$ or $2$. In addition there is no polytropic relationship between the pressure and the density.

Finally, after substituting equation (23) into equations (17) and (18) we obtain
\begin{equation}
\frac{1}{\xi}\frac{d}{d\xi}(\xi\frac{d\theta}{d\xi})+\theta^{n}=0,
\end{equation}
where
\begin{equation}
\theta=[R(y)]^{\frac{1}{n}}, \xi=(n+1)^{-\frac{1}{2}}(\gamma-1)^{\frac{1}{2n}} y, n=1-\nu.
\end{equation}
Equation (24) represents the form of the Lane-Emden equation appropriate for cylindrical polytropes. This equation on the study of the equilibrium of gaseous filaments has been studied (Ostriker 1964). The boundary conditions are
\begin{equation}
\theta(\xi=0)=1, (\frac{d\theta}{d\xi})_{\xi=0}=0.
\end{equation}
It can be easily shown that solutions of the cylindrical Lane-Emden equation satisfying boundary conditions (26) must decrease monotonically from the origin and eventually approach zero. For $n=0$ $(\nu=1)$ or $n=1$ $(\nu=0)$, this equation becomes linear and the solutions can be readily obtained. If $\theta_{n}(\xi)$ reperesnts the required solution, we have
\begin{equation}
\theta_{0}(\xi)=1-\frac{1}{4}\xi^{2}, \theta_{1}(\xi)=J_{0}(\xi).
\end{equation}
We note that in our problem, $\theta_{0}(\xi)$ is not a solution. For other values of $n$, the solutions are not in closed forms. In general, however, Ostriker (1964) obtained a series solution:
\begin{equation}
\theta_{n}(\xi)=1-\frac{1}{(1!)^2 2^2} \xi^{2}+\frac{n}{(2!)^{2} 2^{4}} \xi^{4}-\frac{n(3n-2)}{(3!)^{2} 2^{6}} \xi^{6}+\frac{n(18n^{2}-29n+12)}{(4!)^{2} 2^{8}} \xi^{8}- \ldots .
\end{equation}

We note that the solutions must satisfy the constraint of conservation of mass. In fact, the solutions of the cylindrical Lane-Emden equation are subject to the same homology transformations as are the solutions of the spherical Lane-Emden equation: if $\theta_{n}(\xi)$ is a solution, then $A^{\frac{2}{n-1}}\theta_{n}(A\xi)$ will also be a solution $(n\neq 1)$. Using this transformation, it is possible after finding the solution, then rescale it. Therefore, equation (20) is as follows:
\begin{equation}
\\A^{\frac{2n}{n-1}} \int_{0}^{\xi_{1}} [\theta_{n}(A\xi)]^{n} \xi d \xi= \frac{m}{n+1} (\gamma-1)^{\frac{1}{n}},
\end{equation}
where $\xi_{1}$ is the first zero of $\theta_{n}(\xi)$. All of the solutions $\theta_{n}(\xi)$ have a first zero; i.e., the radii and masses of the filamentary clouds are finite. For mathematical proof of this point see Ostriker (1964). If $n=1$ $(\nu=0)$, then $A \theta_{1}(\xi)$ will be a solution and the equation (29) is replaced by
\begin{equation}
\\A \int_{0}^{\xi_{1}} \theta_{1}(\xi) \xi d \xi= \frac{m}{2} (\gamma-1),
\end{equation}
where $\theta_{1}(\xi)=J_{0}(\xi)$ , so
\begin{equation}
\\A=(\gamma-1)\frac{m}{2\xi_{1} J_{1}(\xi_{1})}.
\end{equation}

Equation (12) shows that for quasi-hydrostatic cylindrically symmetric cooling flows, the density at the center $(\rho_{c})$ increases in proportion to $(t_{0}-t)^{-1}$. This is slower than the growth of density at the center, during gravitational collapse, which is in proportion to $(t_{0}-t)^{-2}$ (Kawachi \&\ Hanawa 1998). We see that in a filamentary cloud $\rho_{c}$ is independent of $\nu$, while in a spherical cloud the central density $\rho_{c}$ depends on $\nu$ and varies in proportion to $(t_{0}-t)^{-\frac{3}{2+\nu}}$ (MMT).

Since the radius of the cloud $(r_{c})$ is defined by the first zero of $\theta_{n}(\xi)$, we can see that $r_{c}$ decreases in proportion to $(t_{0}-t)^{\frac{1}{2}}$, which is independent of $\nu$. MMT showed that in spherical clouds, $r_{c}$ is proportional to $(t_{0}-t)^{\frac{1}{\nu+2}}$ which depends on $\nu$.

Another important difference between cylindrical and spherical cooling flows exists.  In fact, MMT showed that the dynamics of quasi-hydrostatic spherically symmetric cooling  flows depends on $\gamma$, the gas specific ratio. They found that $\gamma=\frac{4}{3}$ is a critical value, i.e., if $\gamma$ is greater (smaller) than $\frac{4}{3}$, then the cloud condenses (expands). But our study showed that in quasi-hydrostatic cylindrically symmetric cooling flows, there is no critical value. In fact, for all values of $\gamma>1$, filamentary clouds undergo radiative condensation.
\section{DISCUSSION AND CONCLUSIONS}
In this paper, we have studied quasi-hydrostatic radiative  cooling flows of  self-gravitating filamentary clouds. A non-Jeans scenario for star formation in these clouds has been investigated. This scenario in spherical clouds was presented, as far as we know, for the first time by MMT.

When the line mass of a filamentary cloud exceeds the critical line mass, the entire filament collapses towards the axis without fragmentation. But a filamentary cloud which is in equilibrium is unstable along its axis and it will fragment. In this paper we suggested another possibility to this view point. A filamentary cloud which is in equilibrium may evolve over a longer time-scale, i.e., due to the cooling of the cloud, quasi-hydrostatic flows can be formed. Considering a simple power-law cooling function, we showed that the filament experiences radiative condensation. These cooling flows can be parameterized by the line mass of the filament.

It is remarkable that the decrement of the radius and the growth of density at the center of the cloud are independent of $\nu$; i.e., we have $r_{c}\propto (t_{0}-t)^{\frac{1}{2}}$ and $\rho_{c}\propto (t_{0}-t)^{-1}$.  We must note that these results strongly depend on the form of the cooling function. But it seems that we can use piece-wise solutions. That is, consider that for some part of the contraction, $A_{\nu}$ and $\nu$ have one set of values, and then join this on to another phase of contraction, when $A_{\nu}$ and $\nu$ have another set of values. What will change is $t_{0}$, but the equations should join with initial and boundary conditions in this way, it seems. If this is true, then we can, in principle, consider a much wider range of cooling functions with the same analytical-type solution. By finding suitable values of $A_{\nu}$  and $\nu$ for the range of densities and temperatures where there is a good approximation, we can find the analytical solution advancing forward from the final state of the previous partial solution. Although we actually do not do this in this paper, using this method we can broaden the applicability of our analysis. Our results would also apply to a turbulent filament, for some range of parameters. In fact, our cooling function in equation (6) is a general function and it might include, in the same qualitative sense, turbulent energy dissipation as well. In other words, one could represent the dissipation rate of turbulent energy as a power law dependent on density and velocity dispersion, and identify the velocity dispersion with the temperature in our equations. There are some nice studies of turbulent dissipation (see, e.g., Mac Low 1999). We suggest that, at least, for some range of parameters, our cooling expression might be used as an approximation to the turbulent cooling rate. Then our results would also apply to a turbulent filament for that parameter range.

Since the behaviours of $r_{c}$ and $\rho_{c}$ obtained based on the similarity solutions, it must be confirmed by a numerical calculation of solutions. We can see that for other forms of cooling function, finding self-similar solution is much more difficult. Thus, numerical simulations will reveal the behaviour of cooling flows in filaments. Accordingly, initial conditions which will lead to the formation of such flows can be determined. We expect that the similarity solutions for describing quasi-hydrostatic cooling flows in filamentary clouds will serve as guidelines for understanding filaments and doing numerical simulations. We must note that our solutions apply only to the early stages of filament contraction, and after the density in the center becomes high enough, perhaps the Jeans-collapse solutions would take over. Even if the Jeans-collapse is eventually important, we have still gained valuable insight into the early stages of filament formation. Since cosmological filaments, diffuse HI filaments, and molecular filaments, all of which show some cooling properties, our solutions might have applications to the formation of these structures.

Each molecular filament typically may contain several distinct clumps which are spaced along that filament. We can consider the filament and the embedded clumps as a self-gravitating system which is in equilibrium (Curry 2000). Theoretical arguments suggest that the mass of each clump is much larger than a typical stellar mass (see, e.g., Fiege \&\ Pudritz 2000). In our scenario, on a longer time-scale the mass and the shape of these clumps have been affected by quasi-hydrostatic cooling flows. It will be interesting to investigate the effects of cooling flows on the clumps.
\\
\\
{\it Acknowledgments}
\\We are grateful to Professor Bruce Elmegreen for his many helpful discussions. MS acknowledges the financial support of Ferdowsi University. This work has made use of NASA's Astrophysical Data System Abstract Service.

\noindent{REFERENCES}
\\Alves, J., Lada, C. J., Lada, E. A., Kenyon, S. J., \&\ Phelps, R. 1998, ApJ, 506, 292
\\Bonnell, I., \&\ Bastien, P. 1991, ApJ, 374, 610
\\Bouquet, S., Feix, M. R., Fijalkow, E., Munier, A. 1985, ApJ, 293, 494
\\Curry, C. L. 2000, ApJ, 541, 831
\\Fiege, J. D., \&\ Pudritz, R. E. 2000, MNRAS, 311, 105
\\Foster, P. N., \&\ Chevalier, R. A. 1993, ApJ, 416, 303
\\Gehman, C. S., Adams, F. C., \&\ Watkins, R. 1996, 472, 673
\\Harjunpaa, P., Kaas, A. A., Carlqvist, P., \&\ Gahm, G. F. 1999, A\&A, 349, 912
\\Houlahan, P., \&\ Scalo, J. M. 1992, ApJ, 393, 172
\\Inutsuka, S., \&\ Miyama, S. M. 1992, ApJ, 388, 392
\\Kawachi, T., \&\ Hanawa, T. 1998, PASJ, 50, 577
\\Larson, R. B. 1969, MNRAS, 145, 271
\\Larson, R. B. 1972, MNRAS, 157, 121
\\Mac Low, M. M. 1999, ApJ, 524, 169
\\Meerson, B., Megged, E., \&\ Tajima, T. 1996, ApJ, 457, 321
\\Myers, P. C. 1991, in IAU Symp. 147, Fragmentation of Molecular Clouds and Star Formation, ed. E. Falgarone \&\ G. Duvert (Dordrecht: Kluwer), 221
\\Nagasawa, M. 1987, Prog. Theor. Phys., 77, 635
\\Nagai, T., Inutsuka, S. I., \&\ Miyama, S. M. 1998, 506, 306
\\Nakamura, F., Hanawa, T., \&\ Nakano, T. 1993, PASJ, 45, 551
\\Nakamura, F., Hanawa, T., \&\ Nakano, T. 1995, ApJ, 444, 770
\\Nakamura, F., \&\ Hanawa, T. 1997, ApJ, 480, 701
\\Nakamura, F., Matsumoto, T., Hanawa, T., \&\ Tomisaka, K. 1999, ApJ, 510, 274
\\Ostriker, J. 1964, ApJ, 140, 1056
\\Penston, M. V. 1969, MNRAS, 144, 425
\\Saigo, K., \&\ Hanawa, T. 1998, ApJ, 493, 342
\\Schneider, S., \&\ Elmegreen, B. 1979, ApJS, 41, 87
\\Semelin, B., Sanchez, N., \&\ de Vega, H. J. 1999, astro-ph/9908073
\\Shu, F. H. 1977, ApJ, 214, 488
\\Spitzer, L., Jr. 1978, Physical Processes in the Interstellar Medium (New York: Wiley)
\\Terebey, S., Shu, F. H., \&\ Cassen, P. 1984, ApJ, 286, 529
\\Wiseman, J. J., \&\ Adams, F. C. 1994, ApJ, 435, 708

\end{document}